\documentclass{article}

\usepackage{PRIMEarxiv}

\usepackage[utf8]{inputenc} 
\usepackage[T1]{fontenc}    
\usepackage{hyperref}       
\usepackage{url}            
\usepackage{booktabs}       
\usepackage{amsfonts}       
\usepackage{nicefrac}       
\usepackage{microtype}      
\usepackage{authblk}
\usepackage{url}
\usepackage[authoryear,round]{natbib}
\usepackage[]{hyperref} 
\hypersetup{hidelinks,
	colorlinks=true,
	allcolors=blue,
	pdfstartview=Fit,
	breaklinks=true}

\usepackage{amsmath, amsfonts, amssymb}

\usepackage[capitalize]{cleveref}

\usepackage{subfigure}

\usepackage{caption}

\usepackage{CJK}

\usepackage{float}

\usepackage{bm} 

\usepackage{fancyhdr}       
\usepackage{graphicx}       
\graphicspath{{media/}}     

\title{Analysis of the background signal in Tianwen-1 MINPA
}

\author[1]{Ziyang Wang}
\author[2]{Bin Miao}
\author[1,4,6]{Yuming Wang}
\author[1,6]{Chenglong Shen}
\author[3,5]{Linggao Kong}
\author[3,7]{Wenya Li}
\author[3,7]{Binbin Tang}
\author[3,8]{Jijie Ma}
\author[3,7,8]{Fuhao Qiao}
\author[3,7,8]{Limin Wang}
\author[3,8,9]{Aibing Zhang}
\author[3,7,8]{Lei Li}

\affil[1]{Deep Space Exploration Laboratory/School of Earth and Space Sciences, University of Science and Technology of China, Hefei, 230026, China}
\affil[2]{Institute of Advanced Technology, University of Science and Technology of China, Hefei, 230026, China}
\affil[3]{National Space Science Center, Chinese Academy of Sciences, Beijing, 100190, China}
				
\affil[4]{Hefei National Laboratory, University of Science and Technology of China, Hefei, 230026, China}
	
\affil[5]{Institute of Science and Technology for Deep Space Exploration, Nanjing University (Suzhou Campus), Suzhou, 215163, China}
	
\affil[6]{CAS Center for Excellence in Comparative Planetology/CAS Key Laboratory of Geospace Environment/Mengcheng National Geophysical Observatory, University of Science and Technology of China, Hefei, 230026, China}
				
\affil[7]{State Key Laboratory of Space Weather, Chinese Academy of Sciences, Beijing, 100190, China}
	
\affil[8]{University of Chinese Academy of Sciences, Beijing, 101408, China}
	
\affil[9]{Beijing Key Laboratory of Space Environment Exploration, Beijing, 100190, China}

\begin{document}
\maketitle

\begin{abstract}
Since November 2021, Tianwen-1 started its scientific instrument Mars Ion and Neutral Particle Analyzer (MINPA) to detect the particles in the Martian space. To evaluate the reliability of the plasma parameters from the MINPA measurements, in this study, we analyze and reduce the background signal (or noise) appearing in the MINPA data, and then calculate the plasma moments based on the noise-reduced data. It is found that the velocity from MINPA is highly correlated with that from the Solar Wind Ion Analyzer (SWIA) onboard the MAVEN spacecraft, indicating good reliability, and the temperature is also correlated with the SWIA data, although it is underestimated and has more scatter. However, due to the limited $2\pi$ field of view (FOV), it's impossible for MINPA to observe the ions in all directions, which makes the number density and the thermal pressure highly underestimated compared to the SWIA data. For these moments, a more complicated procedure that fully takes into account the limited FOV is required to obtain their reliable values. In addition, we perform a detailed analysis of the noise source and find that the noise comes from the electronic noise in the circuits of MINPA. Based on this study, we may conclude that MINPA is in normal operating condition and could provide reliable plasma parameters by taking some further procedures. The analysis of the noise source can also provide a reference for future instrument design.
\end{abstract}


\section{Introduction}
\begin{figure}[H]
	\centering
	\includegraphics[width=0.9\linewidth]{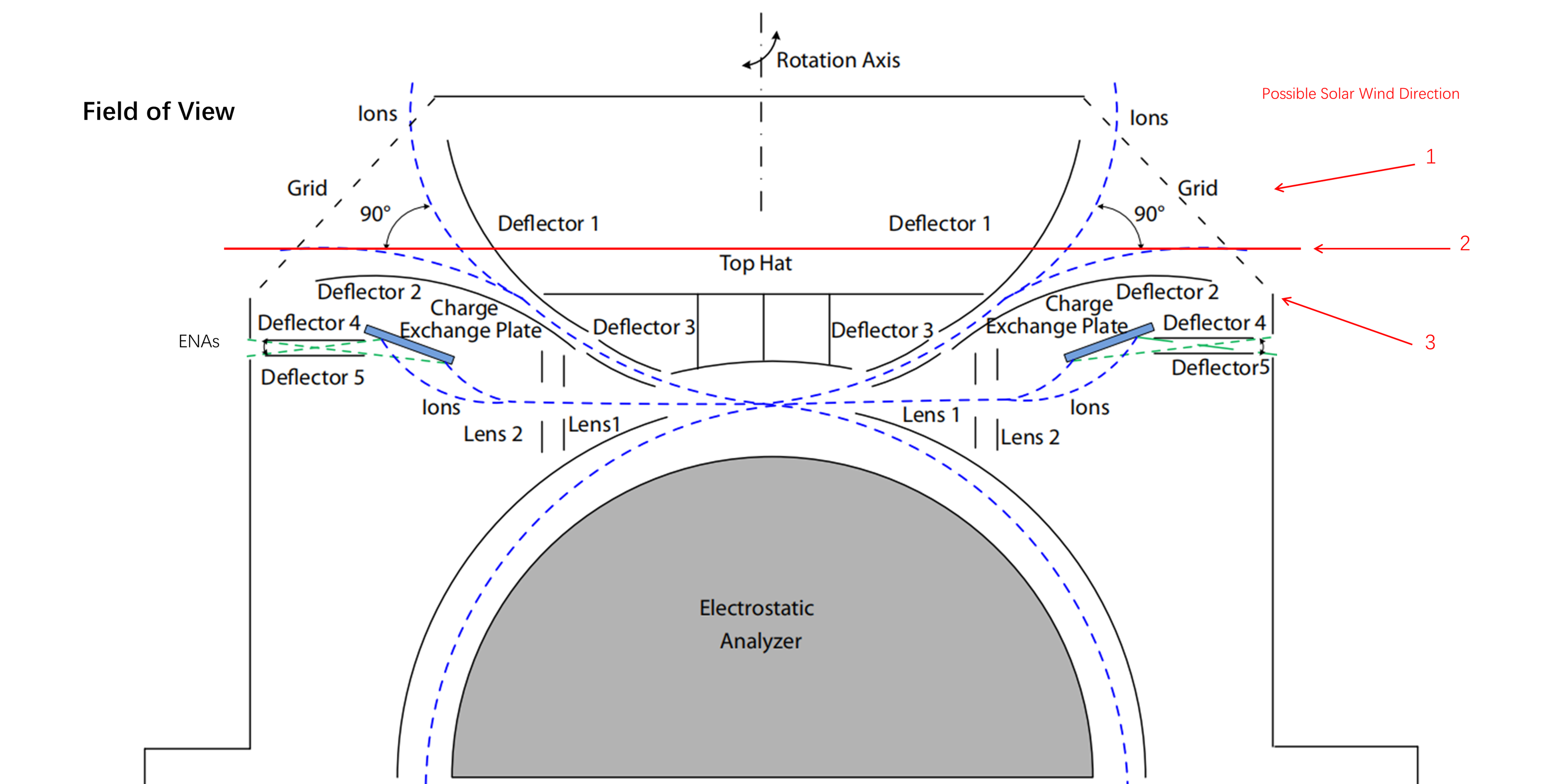}\\
	\caption{Schematic of MINPA's $2\pi$ field of view (FOV), taken from \citet{Kong2020}. The trajectories of the ions and ENAs are shown as blue and green dashed lines, respectively. During most of the time around Mars, the lower surface of the MINPA FOV (the equatorial FOV) is parallel to the x-axis of the MSO coordinate system, i.e. the direction of the Sun-Mars line, which is indicated by the horizontal red line in the figure. The three red arrows indicate possible directions of the solar wind, which will be discussed in ~\cref{noise-reduced data}.}
	\label{fov} 
\end{figure}

The Mars Ion and Neutral Particle Analyzer (MINPA) \citep{Kong2020} is a ion and neutral particle detector onboard China's Mars Exploration Mission (Tianwen-1) orbiter \citep{Wan2020}, which is launched in 2020. MINPA has two units: the ion and ENAs units. These two units are integrated into one sensor head and share a common top-hat type electrostatic analyzer (ESA), a time-of-flight (TOF) mass spectrometer, a micro-channel plate (MCP) sensor, and a electronic unit.  ~\cref{fov} provides a schematic of MINPA, taken from \citet{Kong2020}. The sensor head consists of one ion measurement channel and one ENAs measurement channel, each with its own entrance window. The symmetry axis, the field of view, the deflection system for trajectory adjusting (deflector electrodes) are also illustrated in ~\cref{fov}. The scientific goal of MINPA is to explore the interaction between the solar wind and the neutral and charged particles in the Martian space. As the Tianwen-1 orbiter is designed to cross the interplanetary space, the Martian magnetosheath, and the Martian induced magnetosphere, MINPA is able to continuously observe ions and neutral particles in different regions. 

In-situ low energy particle detectors, such as MINPA, can provide a direct measurement of the particle counts, which can be further converted into energy flux and distribution function \citep{Paschmann1998}. Then the plasma moments such as density, bulk velocity, pressure and temperature can be calculated from the measured distribution function. The background signal or noise counts in such instruments have been analyzed by \citet{Nicolaou2023}, who pointed out that these noise counts, distributed in different energy channels, can change the shape of the measured distribution function and therefore significantly affect the moments calculation. In particular, the solar wind bulk velocity may be underestimated due to the high measured distribution function at low energy channels. Since the solar wind has a narrow distribution function, any high signal in the low-energy band of the distribution function can introduce bias into the moments calculations. Therefore, noise reduction is essential to obtain reliable plasma moments. 

In November 2021, the Tianwen-1 spacecraft released its lander and rover, and the scientific payloads onboard orbiter started their scientific observation mission in the Martian space. Before then, several studies have been done on MINPA observations during the cruise phase of Tianwen-1 in interplanetary space \citep{Zhang2022,Fan2022}. In particular, \citet{Zhang2022} evaluated the blocking effect due to the lander capsule and showed the solar wind plasma moments during a stream interaction region (SIR) event. While \citet{Fan2022} suggested that the contamination of the detector noise is mainly centered around several azimuth angles and therefore simply treated the flux signal at these azimuth angles as noise and subtract it from the original data.

However, considering that MINPA is in different modes of operation during the cruise phase and in orbit around Mars, the noise reduction method used in previous studies to deal with interplanetary data may not be suitable to data in orbit around Mars. Thus, the noise analysis and reduction is urgently needed for the data observed in the Martian space, though the preliminary scientific data in the Martian environment have been released. In this study, we introduce and then justify our preliminary method of noise reduction in ~\cref{data and methodology}. The noise-reduced data and their comparison with the data from the Solar Wind Ion Analyzer (SWIA) onboard MAVEN \citep{Halekas2015} are presented in ~\cref{noise-reduced data}, and the detailed analysis of the noise source is performed in ~\cref{the analysis of noise source}. In ~\cref{conclusion}, we conclude the study of the paper.

\section{Data and methodology} \label{data and methodology}

\subsection{Data} \label{data}

\begin{figure}[H]
	\centering
	\includegraphics[width=0.9\linewidth]{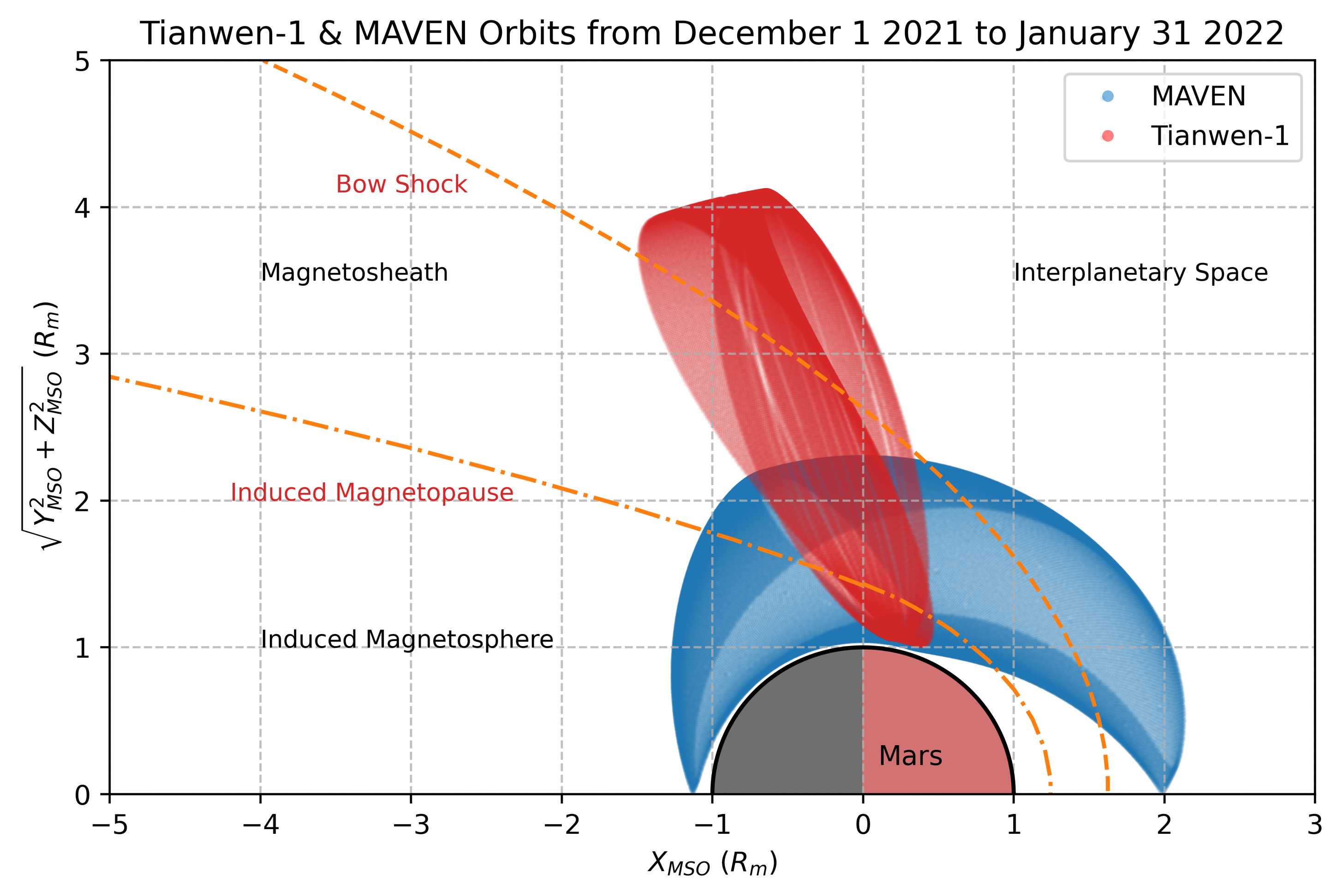}\\
	\caption{Orbits of Tianwen-1 (red) and MAVEN (blue) from December 1, 2021 to January 31, 2022. The orbits are shown in the MSO coordinate system in cylindrical form. The bow shock and induced magnetopause models are given by \citet{Trotignon2006}}
	\label{orbit} 
\end{figure}

MINPA is designed to detect the ions and energetic neutral atoms (ENAs) in the Martian environment, which has three main observational modes: the default mode, the magnetotail mode (near apoapsis), and the ionosphere mode (near periapsis) \citep{Kong2020}. The default mode operates between the periapsis and apoapsis of Mars, which is mainly intended to detect the solar wind and the magnetosheath. It has 40 energy channels (W) covering the energy range from 2.8\,eV to 25.3\,keV, 8 mass channels (M) ranging from 1 to 70 atomic mass units (amu) that can distinguish ion species such as H$^+$, He$^+$, O$^+$, O$_2^+$, CO$_2^+$, 4 elevation angle channels (E) and 16 azimuth angle channels (A) covering the $2\pi$ field of view (FOV) with a time resolution of 16\,s. The magnetotail mode operates for approximately 30 min near apoapsis and generates data with $64W \times 8M \times 16A \times 4E$ with enhanced energy and temporal resolution. In the ionosphere mode operating near periapsis, MINPA sums up the $4\times16$ angular channels and provides data with $48W \times 32M$, sweeping ion energies from 2.8\,eV to 3.4\,keV with a time resolution of 4\,s. In the following analysis, we only focus on the default mode.

To verify the reliability of noise-reduced MINPA data, we use the plasma moments observed by MAVEN SWIA, which does not discriminate between ion species, but can detect protons in the energy range of 25\,eV to 25\,keV with a good spatial resolution \citep{Halekas2015}. It has a time resolution of 4 seconds and can provide H$^+$ moment data such as number density and velocity of protons.

We will use the data from December 1, 2021 to January 31, 2022 to analyze and remove noise in this study. The orbits of Tianwen-1 and MAVEN are shown in ~\cref{orbit} in the Mars Solar Orbital (MSO) coordinate system. As shown in the figure, the orbits of both satellites cover three regions surrounding Mars during this period. As we know, the solar wind is a large-scale homogeneous structure, which means that when both satellites are in the solar wind near Mars, their observations should be comparable. Thus, we can check the reliability of the noise-reduced MINPA data by comparing them with the SWIA data. The similar validation method was also used in \citet{Fan2022}. They use the time shift method in \citet{opitz2009,opitz2010}, which suggests that the solar wind parameters from multi-spacecraft in the inner heliosphere can be shifted and scaled by assuming that the velocity is a constant and the density decreases by $R^{-2}$, where $R$ is the heliocentric redial distance, when their longitudinal separation is less than 65$^{\circ}$. Considering that the Tianwen-1 and the MAVEN are both near the Mars during the period of our study, we believe that this method of comparison is practical.

\subsection{Methodology of the noise reduction} 

\begin{figure}[H]
	\centering
	\includegraphics[width=\linewidth]{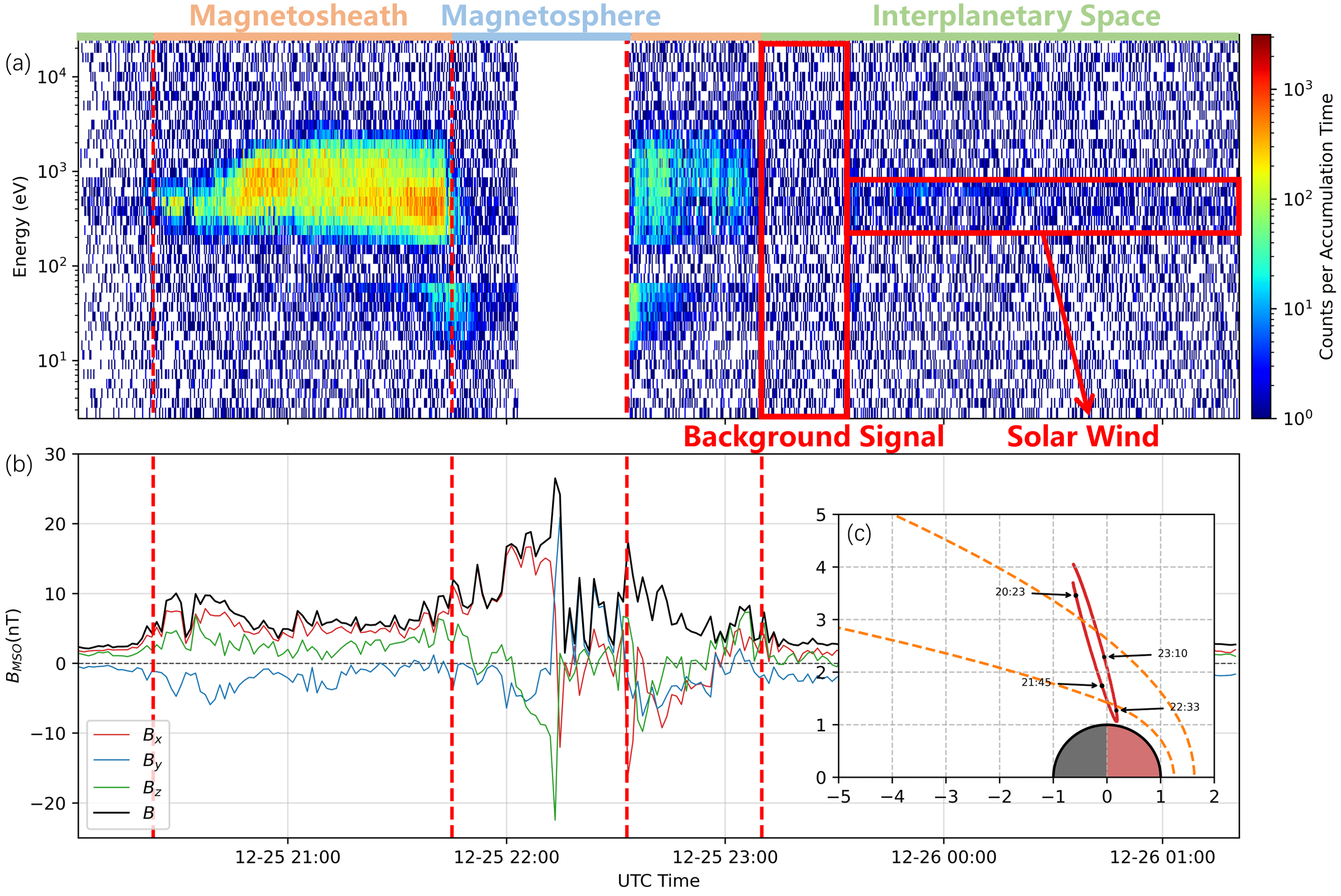}\\
	\caption{(a) The H$^+$ energy spectrum observed by MINPA default mode during one orbital period from December 25 20:02 UT to December 26 01:20 UT, which is summed up for all directional channels. The colors indicate the measured counts, sampled in a 0.019 s accumulation period for each channel. The red dashed lines and the color bars above the figure indicate different regions. The red boxes and testes are used to indicate whether or not the solar wind was observed during the time period. During the empty period, MINPA switches to another observation mode. Due to the voltage supply error near 100 eV, the observation of ions is suppressed, resulting in a sudden drop in the counts at around 100 eV. (b) The magnetic field observed by the MOMAG magnetometer onboard Tianwen-1 \citep{Zou2023}, and the red dashed lines have the meaning as above. (c) The orbit of Tianwen-1 during this period with the same format as ~\cref{orbit}. The black dots correspond to the times of the red dashed lines when Tianwen-1 passes the interfaces of different regions. }
	\label{background} 
\end{figure}

~\cref{background} (a) shows an example of MINPA observations in one orbital period from December 25 20:02 UT to December 26 01:20 UT. To investigate the observational characteristics of different regions in the Martian space, we first distinguish the regions that Tianwen-1 traveled through as described below.

In the interplanetary space, energy spectrum has a narrow span and magnetic field remains roughly constant because the solar wind is a cold and large-scale homogeneous plasma. In the magnetosheath, the energy spectrum is broadened and there is a sudden increase in the magnetic field due to the bow shock in front of the magnetosheath. In the induced magnetosphere, the energy and counts of H$^+$ decrease in the spectrum, and the magnetic field undergoes further enhancement and fluctuation. The red dashed lines in ~\cref{background} (a) and (b) correspond to the interfaces identified according to the above criteria. As can be seen in these panels, the positions of the interface features in the energy spectrum and the magnetic field show good agreement, which is indicated by the red dashed lines. 

On the other hand, we also use the bow shock and induced magnetopause models given by \citet{Trotignon2006} as a complement. As shown in ~\cref{background}(c), the black dots corresponding to the red dashed lines locate near the interfaces. Considering that the interface model reflects a long-term average situation, we believe that the regions identified according to the energy spectrum and the magnetic field can reflect the real situation.

After determining regions, we find that when the satellite is in the interplanetary space, a relatively clear solar wind count signal (at least 1 order of magnitude higher than background counts) can be observed during certain periods, such as around December 26 00:00 UT in ~\cref{background} (a) (near $10^3$\,eV, indicated by red boxes with text "Solar Wind"). While in some other time periods like around December 25 23:10 UT (indicated by red boxes with text "Background Signal"), almost no solar wind count signal but pure background noise was observed. Both of the above situations are marked with red boxes in ~\cref{background} (a). In fact, the clearness of the solar wind signal is affected by the hemispheric viewing field of the MINPA and the attitude of Tianwen-1, which will be described in detail in ~\cref{noise-reduced data}. As mentioned above, the presence of background noise will interfere with the calculation of moments for plasma flows \citep{Paschmann1998}, especially when the level of noise in our data is comparable to that of a weak solar wind signal. Thus, we need to remove the background noise from the signal before we can formally calculate the plasma moments.

The method we use to determine the background noise is to select a series of cases like ~\cref{background} (a) around December 25 23:10 UT, in which the solar wind signal is not observed at all, each case lasting five to ten minutes. For each case, we calculate the average value of the non-zero count on each channel over the corresponding period and treat it as the background noise of the case. Then we calculate the average background noise of these cases and use it in the next step. This treatment can also be seen in some previous studies \citep{Nenon2019, Franz2006, Nicolaou2023}. 

\begin{figure}[H]
	\centering
	\includegraphics[width=\linewidth]{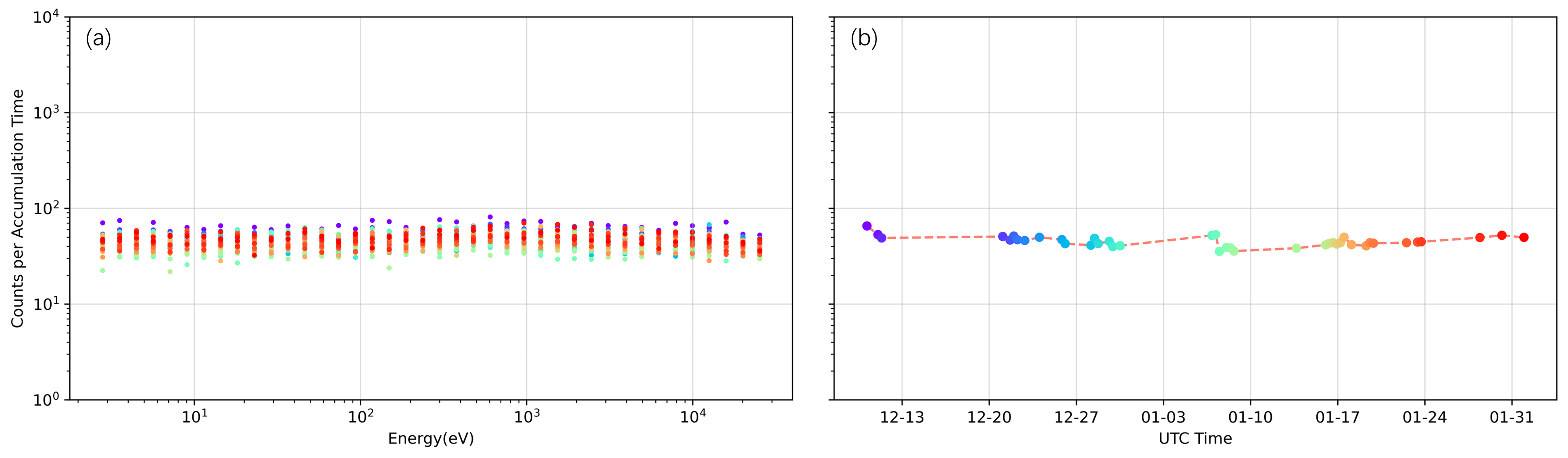}\\
	\caption{Background signal counts determined by forty cases during the period from December 1 2021 to January 31 2022, in which the solar wind signal can be neglected. Different colors denote cases selected in different time periods. (a) Background noise of MINPA default mode. (b) Average noise counts of all energy channels for each case over time, with colors corresponding to (a).}
	\label{background_noise_comparison} 
\end{figure}

During the period from December 1 2021 to January 31 2022, the Tianwen-1 had been in orbit around Mars for about two hundred cycles, and forty cases are analyzed and shown in ~\cref{background_noise_comparison}. As can be seen in ~\cref{background_noise_comparison} (a), the background noise of these cases shows great consistency and is spread over a relatively narrow count range. Moreover, ~\cref{background noise comparison} (b) also shows that the average noise of all energy channels is almost constant for each case during this period. Thus, we believe that this count range is narrow enough that the background noise can be considered invariant with time, and the average background noise calculated over all cases can represent the noise level of MINPA during this time period. Then we subtract the average background noise from all the count data during this period to obtain the noise-reduced data.

\section{Noise-reduced data} \label{noise-reduced data}

\begin{figure}[H]
	\centering
	\includegraphics[width=\linewidth]{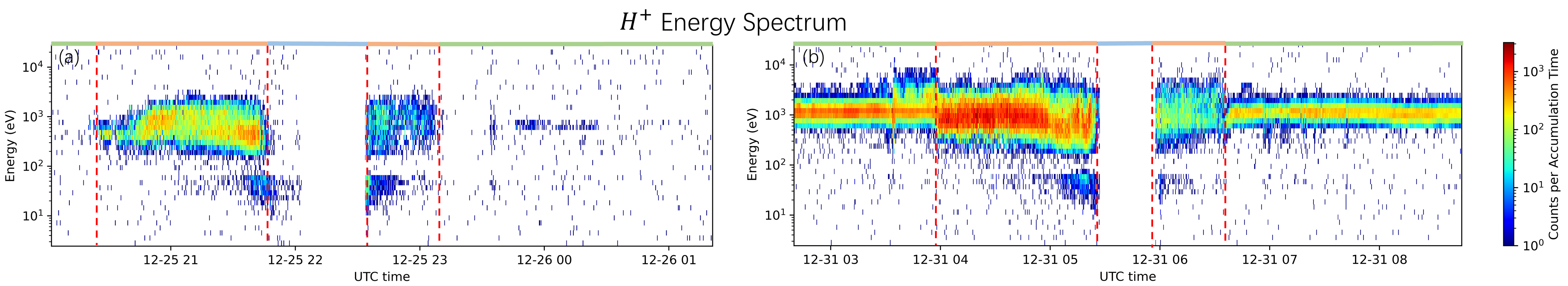}\\
	\caption{H$^+$ energy spectrum after noise reduction. The left figure shows the situation when the solar wind signal is weak, while the right figure corresponds to a clear solar wind signal. The red dashed lines and the color bars above the figure have the same meaning as in ~\cref{background} (a)}
	\label{after_noise_reduction} 
\end{figure}

The energy spectrum after noise reduction is shown in ~\cref{after_noise_reduction}. The left panel has the same time period as ~\cref{background} (a) and the right panel provide a case in which a clear solar wind signal is observed from December 31 02:40 UT to December 31 08:40 UT. After applying the noise reduction method, we can see that the background noise is significantly removed by comparing ~\cref{background} (a) and ~\cref{after noise reduction} (a), while the magnetosheath signal and the weak solar signal is preserved. These results are the basis for the subsequent calculation of the plasma moments.

\begin{figure}[H]
	\centering
	\includegraphics[width=\linewidth]{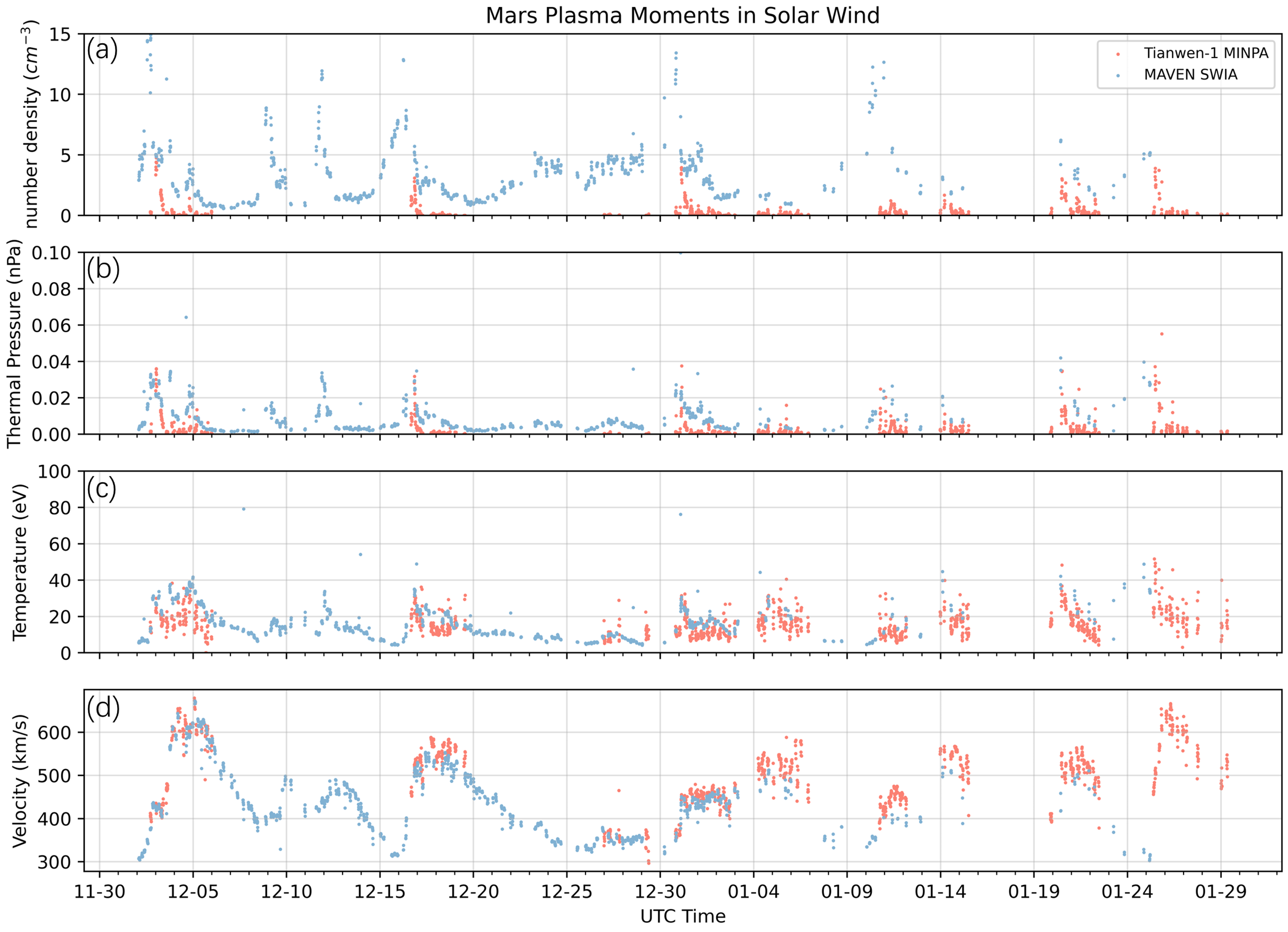}\\
	\caption{Comparison of the proton plasma moments in the solar wind between Tianwen-1 MINPA (red) and MAVEN SWIA (blue) from December 1, 2021 to January 31, 2022, including number density, velocity, thermal pressure and temperature.}
	\label{solar_wind_moments} 
\end{figure}

There are two methods to derive the plasma moments: one is the integral method and the other is the fitting method (see details in Appendix A). We apply both methods to the noise-reduced data and find that the fitting method is more robust to the data here in deriving velocity and temperature. Since the integral method requires integration over the entire phase space, and the MINPA data sometimes suffer from the limited $2\pi$ FOV, the integral method will fail if the observation of the ion distribution is limited in several directions. Therefore, we use the fitting method to calculate velocity and temperature, while the integral method is used to calculate number density and thermal pressure directly. As mentioned in ~\cref{data}, the solar wind is a large-scale homogeneous structure, and therefore we can compare the derived plasma moments of MINPA and SWIA when both satellites are in the solar wind. Therefore, we focus our following comparison on the solar wind region. The results of the plasma moments from December 1, 2021 to January 31, 2022 are shown in ~\cref{solar_wind_moments}. We also show the correlation of moments in ~\cref{solar_wind_correlation} and compare the statistical distributions of velocity and temperature between MINPA and SWIA in ~\cref{solar_wind_statistics}.

It should be admitted that not all derived plasma moments are in agreement with SWIA. In general, the derived parameters related to number density (number density and thermal pressure) are highly underestimated, and the temperature is slightly underestimated, while the solar wind velocity shows a good agreement, which will be discussed in the following sections.

\subsection{Number density and thermal pressure}

The underestimation of number density and thermal pressure is related to the limited $2\pi$ FOV of MINPA and the attitude of Tianwen-1. For most of the time around Mars, the equatorial FOV is kept roughly parallel to the Sun-Mars line, which is indicated by the horizontal red line in ~\cref{fov}. Since the solar wind flow is known to be mainly along the radial direction, the instrument's FOV should be able to capture solar wind ions. However, any disturbance may cause the solar wind deviating from the x-axis and the limited $2\pi$ FOV makes it impossible for MINPA to observe ions in all directions. If the deviated solar wind direction is outside the FOV (as indicated by arrow 3 in ~\cref{fov}), the instrument may partially or even completely miss the main component of the solar wind population. Furthermore, we know that the derived value of number density, velocity, and temperature depend on the height, peak position, and width of the distribution function, respectively. So the incomplete capture of ions will significantly reduces the height of the distribution function, resulting in the underestimated number density.

As we can see in ~\cref{after_noise_reduction} (a), there is almost no count signal when the satellite located in the solar wind because the FOV of MINPA almost completely missed the solar wind ions, which will result in the underestimated number density. In contrast, when the MINPA FOV captures the main component of solar wind population, a clear solar wind signal is observed, as shown in ~\cref{after_noise_reduction} (b). The reason for the underestimated thermal pressure is similar, since the thermal pressure in the fitting method is related to the number density as $p=n \kappa T$. Moreover, from the scatter plots in ~\cref{solar_wind_correlation} (a, b), we may find that there is no strong correlation in the number density and pressure between MINPA and SWIA. It implies that we cannot derive the reliable values of the number density and pressure from MINPA instrument based on any simple correction procedure, unless the limited FOV is fully considered and simulated during the correction procedure.

\subsection{Temperature}

\begin{figure}[H]
	\centering
	\includegraphics[width=\linewidth]{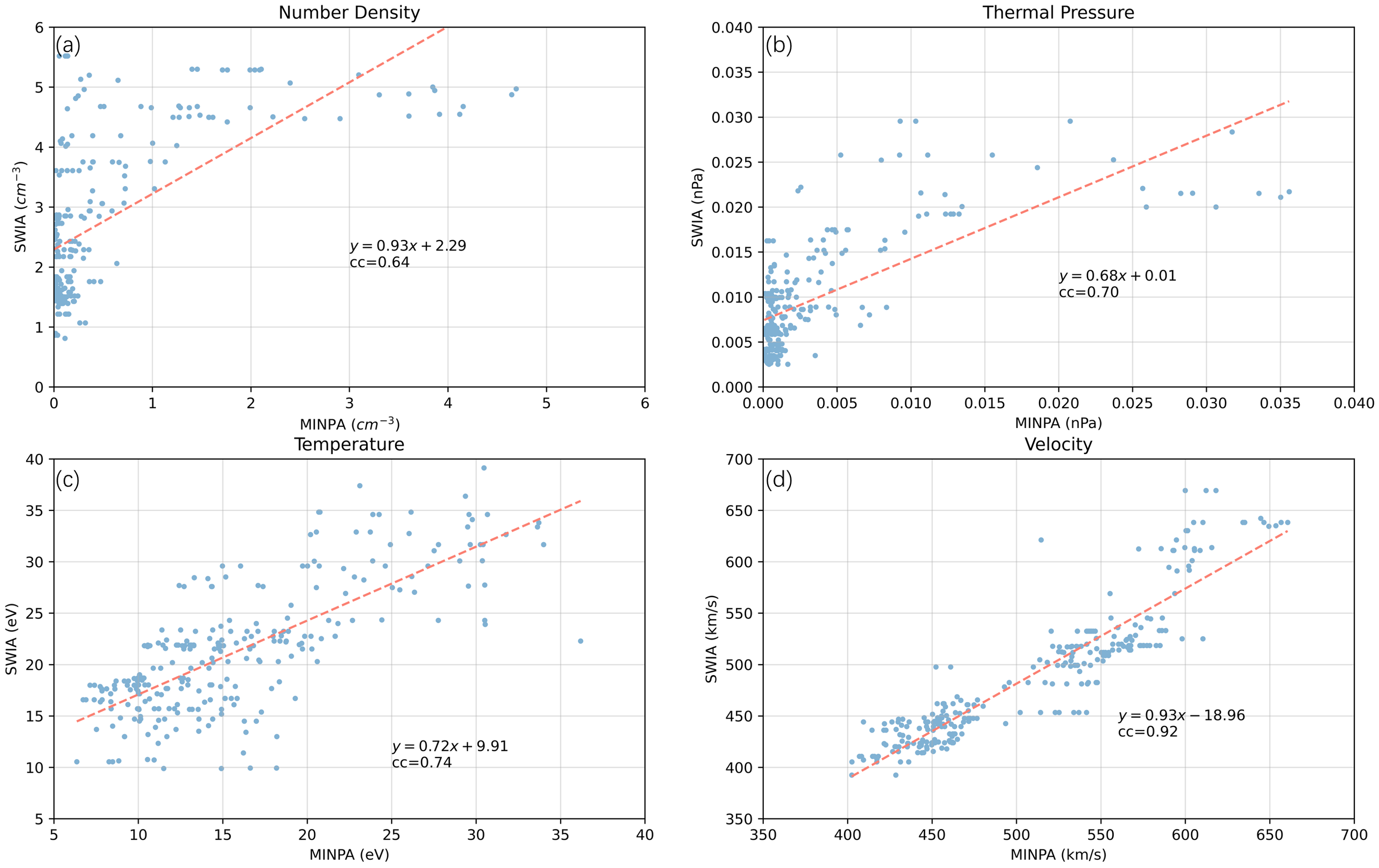}\\
	\caption{Correlation between moments from MINPA and SWIA with linear fitting, and the texts present the fitting expressions and Pearson's correlation coefficients (cc). Each point represents a MINPA data with the corresponding nearest SWIA data when Tianwen-1 and MAVEN are in the solar wind at the same time (i.e. not including all the data in ~\cref{solar wind moments}).}
	\label{solar_wind_correlation} 
\end{figure}

\begin{figure}[H]
	\centering
	\includegraphics[width=\linewidth]{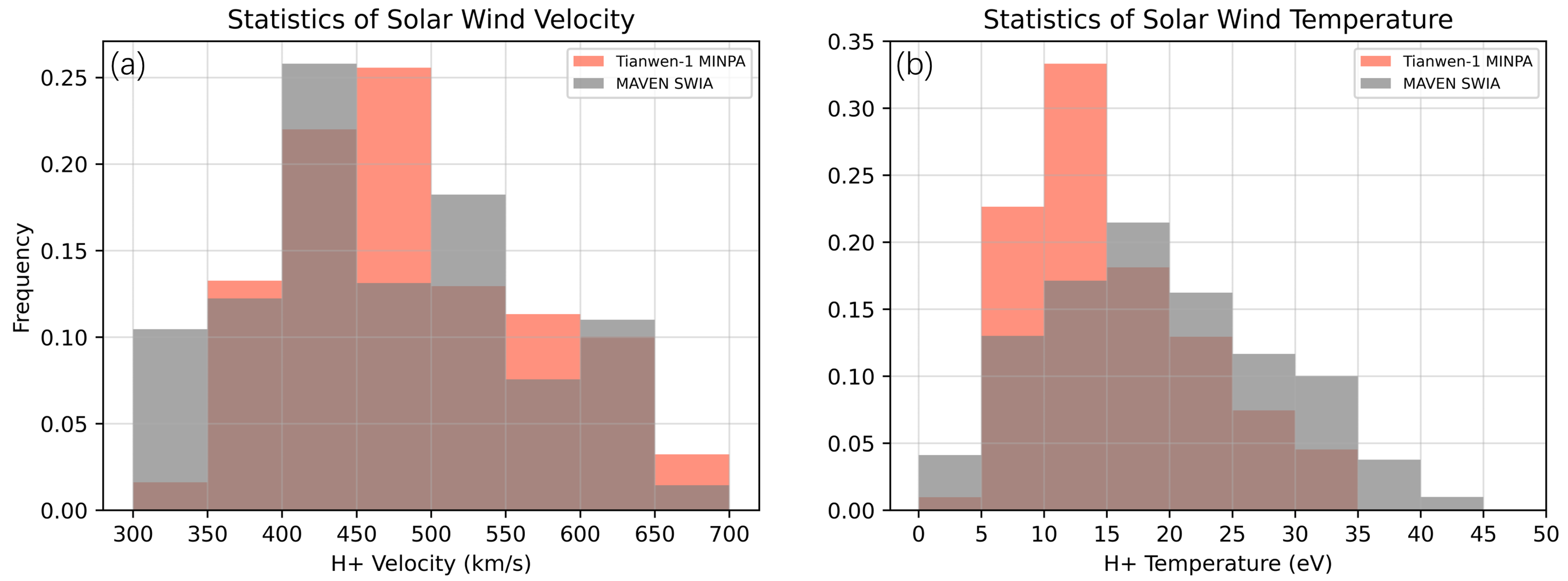}\\
	\caption{Statistics of velocity and temperature between MINPA and SWIA, with the same data set as ~\cref{solar wind correlation}.}
	\label{solar_wind_statistics} 
\end{figure}

The temperature observed by MINPA also shows a slight underestimation compared to that observed by SWIA. But different from the number density and pressure, its value shows a much better correlation with that from SWIA as displayed in ~\cref{solar_wind_correlation} (c). According to the linear fitting, the temperature value from MINPA may be able to be corrected by using $T_{corrected} = 0.72 \times T_{original} + 9.91$ as the first order approximation if we consider the moments from the SWIA to be reliable, though the cc value is only 0.74 and scattering is also large. As the temperature show more credible correlation between the two instruments, we further show the distribution in ~\cref{solar_wind_statistics} (b), which suggests that the temperature from MINPA is underestimated by about 21 percent compared to MAVEN on average. A possible explanation is that the signal strength on both sides of the peak of the distribution function is comparable to the noise level, so that almost only the signals close to the peak remain after noise reduction, which can lead to an underestimation of the width of the distribution function and therefore the temperature.

\subsection{Solar wind velocity}

Despite the underestimated height and width of the distribution function, we can obtain a relatively reliable velocity as long as the distribution of the captured ions is similar to that of the original solar wind. As can be seen in ~\cref{solar wind moments,solar wind correlation,solar wind statistics}, the solar wind velocity from MINPA and SWIA show a good agreement with the cc value as high as 0.92, since the peak position of the distribution function can be accurately estimated. Based on the linear fitting in ~\cref{solar wind correlation} (d), the value of solar wind velocity from MINPA can be related to that from SWIA with the empirical function $v_{SWIA} = 0.93 \times v_{MINPA} - 18.96$, suggesting that the MINPA values are slightly overestimated. We also show the distributions of velocity in ~\cref{solar wind statistics} (a), which further suggests that the average deviation of velocity from MINPA and SWIA is within 3 percent, indicating a good consistency.

\section{The analysis of noise source} \label{the analysis of noise source}

The noise contamination in MINPA observations can come from a variety of sources, including electronic noise in the detector's electrical circuits, ultraviolet (UV) radiation from the Sun, Galactic Cosmic Ray (GCR) on a long-term basis, Solar Energetic Proton (SEP) on a transient basis, and the secondary electrons generated from the detector's inner surface by the collision of UV photons and energetic particles \citep{Wuest2007,Franz2006,Paschmann1998,Nenon2019}. 

Electronic noise presenting in electrical circuits acts as stochastic background noise. The noise-induced counts are uniformly distributed across all energy levels of ESA and mass channels of TOF. This type of noise is likely generated after the MCP sensors and is independent of the electric fields applied to ESA and TOF for charged particle selection.

Fine serration and black coating have been used in UV photon suppression \citep{Saito2017, Gershman2016, Gilbert2014}, but UV photons can enter the ESA through the entrance of ENAs. Because ENAs must be ionized without significant kinetic energy change, the entrance system is designed to be nearly straight, as shown in ~\cref{fov}. This type of flat trajectory entrance system cannot efficiently block UV photons. Fortunately, most of the UV photons can be absorbed by the charge exchange plate behind the entrance (as shown in ~\cref{fov}) and only a few photons can enter and hit the ESA, so the UV contamination is limited and lasts only a few tens of minutes. In addition, considering that the source of UV photons is the Sun, the UV contamination is directional and mainly distributed on the side of the detector facing the Sun. This makes it easy to identify UV contamination based on its temporal and directional characteristics \citep{Paschmann1998}.

GCR can penetrate space instrumentation and generate counts directly on the MCP. However, unless the MCP is directly struck, the noise level caused by GCR is several orders of magnitude lower compared to other noise sources. \citet{Nenon2019} shows that the instrumental noise due to GCRs is negligible on a long-term basis compared to other noise sources such as energetic particles or electronic noise. Thus, this background has almost no effect on the long-term average of the counts, which leads us to exclude this source first. On the other hand, the penetrating particle noise is only important when there is a high flux of energetic particles, such as during SEP events. Therefore, this noise is usually transient in nature. 

The secondary electrons generated by UV, GCR and SEP may also contribute to the noise counts. However, the electric field controlled by ESA and TOF may suppress a part of these electrons, and the criterion of TOF signal collection may discriminate another part of these electrons. In the following sections, we will analyze the MINPA noise in detail according to the characteristics of the different noise sources mentioned above.

\begin{figure}[H]
	\centering
	\includegraphics[width=\linewidth]{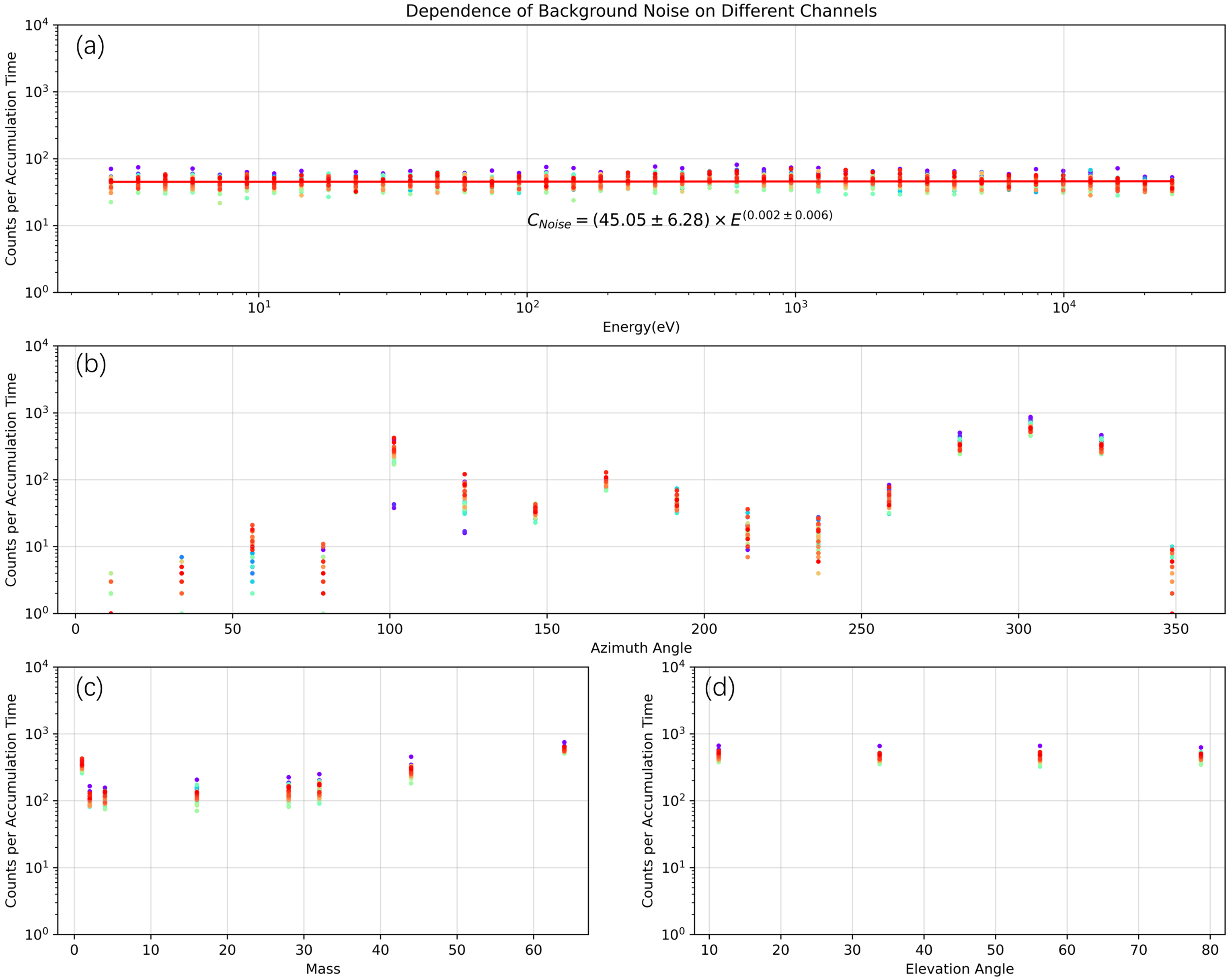}
	\caption{Dependence of the background noise on different channels, including (a) ion energies, (b) azimuth angles, (c) ion masses, (d) elevation angles. Different colors denote different cases that are the same as those in ~\cref{background noise comparison}.}
	\label{dependence} 
\end{figure}

\subsection{The dependency of noise on different channels}

First, we show the dependence of the background noise on different channels (including ion energies, azimuth angles, ion masses and elevation angles) in ~\cref{dependence}. If the noise is caused by solar UV, then it should be much stronger in the FOV facing the Sun and appear as a transient event. But as can be seen in ~\cref{dependence} (b), the noise dependence on azimuth doesn't show this specific trend. Instead, the background noise seems to higher around three directions, $100^{\circ}$, $170^{\circ}$, $300^{\circ}$, than other directions. In addition, as described above, the background noise is time-stable on the time scale of several months in this study, which is inconsistent with the timescale of several minutes of UV radiation. Therefore, the background is less likely to be generated by solar UV radiation. Similarly, the penetrating particle noise caused by the SEP is often transient in nature, which can also be ruled out. 

From the above analysis, it can be concluded that the background noise is most likely the electronic noise in the electrical circuits of the detector. And under this conclusion, the time stability can be easily explained. Because it's always reasonable to assume that the electronic noise of an instrument will remain at a stable level for a relatively long period of several months in this study. In addition, the similarities between MINPA noise and electronic noise can also be showed as follows.

\subsection{Determining the noise source}

According to the design of MINPA, the analyzer spends the same time on every energy channel in an accumulation period. Thus, the electronic noise in the ion counts should be at the same level for each energy channel. Based on the above analysis, we examine the energy spectrum of the noise counts by fitting them with a power law function. The fitting result is shown in ~\cref{dependence} (a). The power index of the fitted curve is close to zero with a small standard deviation, suggesting that the noise is independent on the energy. Thus, the background noise of MINPA is probably the electronic noise of the instrument.

\begin{figure}[H]
	\centering
	\includegraphics[width=\linewidth]{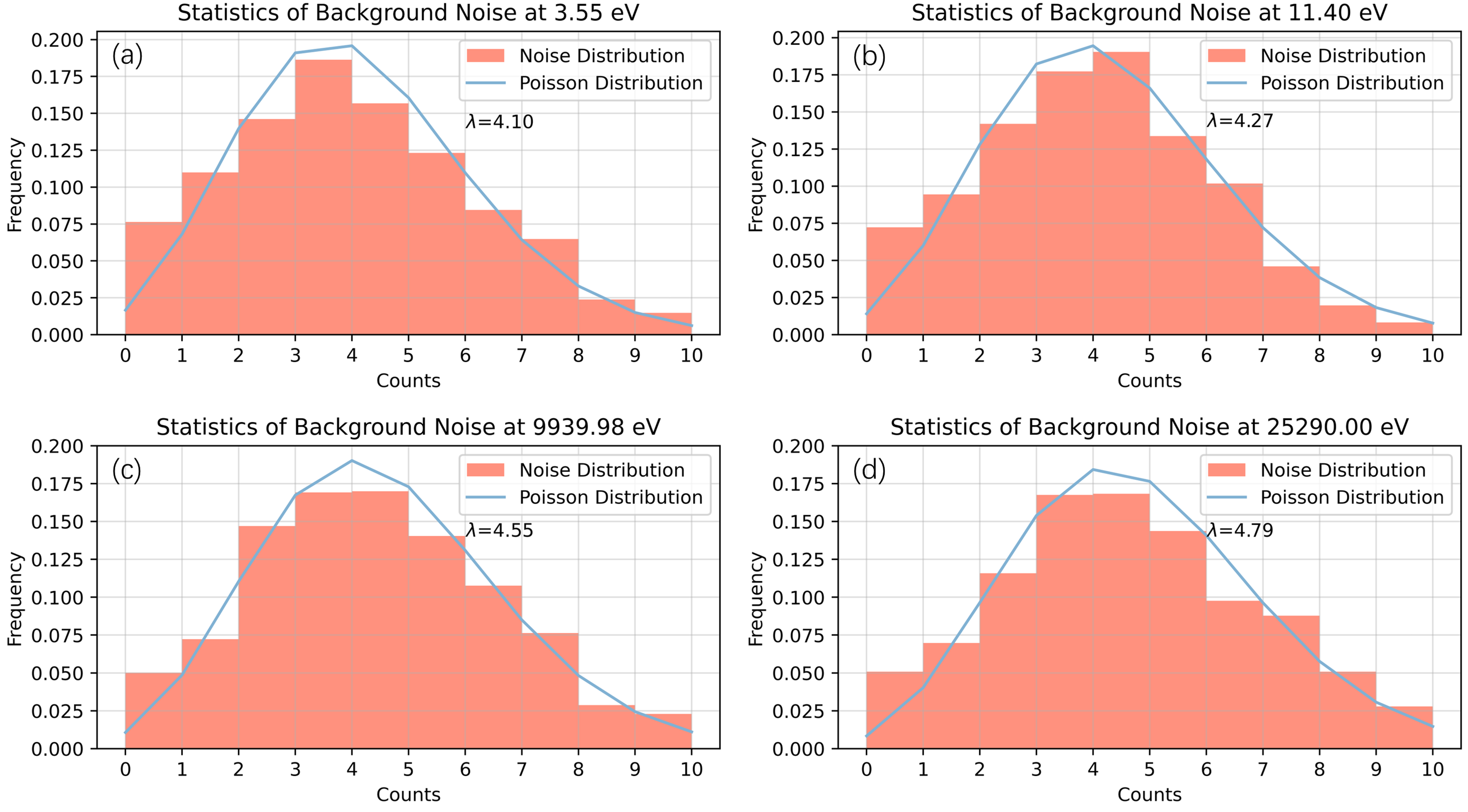}\\
	\caption{The red histograms denote the statistical distributions of the background noise at different energy channels from December 29 15:00 UT to December 30 18:00 UT, and the blue lines denote the corresponding Poisson distributions with fitted parameters $\lambda$, whose values are given in the figure.}
	\label{noise_statistics} 
\end{figure}

The assumption of uniform noise is also found in several studies. For example, \citet{Franz2006} and \citet{Nicolaou2023} both suggested that the signals on higher or lower energy channels, where we do not expect to measure valid ion signals, can be treated as background noise and be used as a representative noise level for all energy channels. Thus, they subtracted the noise signal from the original signal for all energy channels. Obviously, this subtraction procedure requires the assumption of uniform noise across energies. While our study confirms this assumption by showing the energy spectrum of the background noise in ~\cref{dependence} (a). 

On the other hand, we will show another example to further demonstrate that the background is probably electronic noise. For random events, the Poisson distribution is widely used to describe the probability distribution of the events occurring in a time period. Thus, if the background noise is electronic noise, it should obey the Poisson distribution. This treatment can also be found in \citet{Nicolaou2023}. In addition, as described above, the signals on higher or lower energy channels can be treated as background noise, so we show the statistical distributions of background noise on the specific energy channels from December 29 15:00 UT to December 30 18:00 UT in ~\cref{noise_statistics}. The statistical distribution and the Poisson distribution fit well, providing another piece of evidence. We also note that the parameter value of the Poisson distribution (i.e., the mean count of the background noise) for each energy channel is close to each other, which supports our earlier statement that the electronic noise is at the same level for each energy channel.

\subsection{Location of electronic noise generation in MINPA}

Referring to ~\cref{fov} and the complete schematic given by \citet{Kong2020}, it can be seen that the ESA section contains only simple deflection voltages and does not contain any electrical circuit. Thus, the electronic noise can only be generated after the ESA, such as the MCP sensors in the TOF. Taking it a step further, \citet{Wuest2007} suggests that TOF system inside the analyzer is inherently less susceptible to noise because of the coincidence requirement for ion detection. A rough explanation can be given as follows: in order to detect a valid ion signal, two or three electronic signals must be triggered within a given time window. For example, to be identified as a valid signal, an ion must trigger a 'start', a 'stop', and perhaps a 'position' or 'energy' detector. Therefore, the random electronic noise may only result in a signal on a single detector and will not be identified as a valid signal, nor will it create a background noise.

Besides, in ~\cref{dependence}, the clear dependence of the background noise on azimuth angles and ion masses should also be noted, showing the possibility that the background noise is generated during the process of ion channel identification. All of the above analysis leads us to conclude that the background noise is most likely coming from the electronics unit located at the end of the MINPA.

\section{Conclusions} \label{conclusion}

In this study, we provide a method to estimate and reduce the background noise in MINPA counts data. Then we use the noise-reduced data to calculate plasma moments and compare them with those of MAVEN SWIA, and we also analyze the source of the noise. Our conclusions are listed as follows:

\begin{itemize} 
	\item The background noise is independent on time and energy during the period from December 1 2021 to January 31 2022.
	\item The background noise in the MINPA data is most likely electronic noise, and is generated in the electronics unit located at the end of the MINPA.
	\item The number density and thermal pressure of MINPA are highly underestimated due to the limited $2\pi$ FOV, and therefore there is no correlation in the two parameters between MINPA and SWIA. To obtain their reliable values, a more complicated procedure with the limited FOV fully considered is required.
	\item The temperature observed by MINPA also shows a slight underestimation compared to that observed by SWIA. But different from the number density and pressure, its value shows a much better correlation with that from SWIA and may be able to be corrected by using $T_{corrected} = 0.72 \times T_{original} + 9.91$ as the first order approximation if we consider the moments from the SWIA to be reliable.
	\item  The solar wind velocity from MINPA and SWIA show a good agreement with the cc value as high as 0.94 and can be related with the empirical function $v_{SWIA} = 0.93 \times v_{MINPA} - 18.96$, which suggests that the MINPA values are slightly overestimated.
\end{itemize}

In this study, the background signal in default mode, which mainly operates in the solar wind and magnetosheath, is analyzed. The results suggest that MINPA is in normal operating condition. But the limited FOV has to be considered to derive reliable plasma moments except the velocity. The methods used in this study can be served as a basis for future establishing a pipe line to process the MINPA data. With a joint-observations from Tianwen-1 MINPA and MAVEN SWIA, we are looking forward to the improvement of our understanding of the Martian plasma environment.

\section*{Appendix A: Plasma moments calculation} \label{appendix}

The methodology used to calculate plasma moments in this study is mainly based on \citet{Franz2006}, which has also been widely used to process data from Mars Express \citep{Wei2012,Dubinin2006}, MAVEN \citep{Fan2020,Dubinin2020}, Tianwen-1 MINPA during the cruise phase \citep{Fan2022} or to simulate plasma observations \citep{Nicolaou2023}. And for a general introduction to the calculation of moments, see \citet{Hutchinson2002}, \citet{Wuest2007} and \citet{Paschmann1998}.

We assume that each particle species detected by the analyzer for an accumulation period can be described by a distribution function $f(\bm{v})$ in velocity space. And the relation between the distribution function $f(\bm{v})$ and the counts $C(E, \theta, \phi)$ for particles of energy (E) detected in the direction indicated by the azimuth angle $\phi$ and the elevation angle $\theta$ is: \citep{Wuest2007}

\begin{equation} \label{c to f}
	\begin{split}
		f(E, \theta, \phi) &\ = \frac{J(E, \theta, \phi)m^2}{2E^2} = \frac{2J(E, \theta, \phi)}{v^4} \\
		J(E, \theta, \phi) &\ = \frac{C(E, \theta, \phi)}{G \Delta T \frac{\Delta E}{E}} 
	\end{split}
\end{equation}

Where $J(E, \theta, \phi)$ is the energy flux, $G$ is the geometric factor, $\Delta T$ is the accumulation period for each channel, $\Delta E$ is the energy span at energy $E$. Then the plasma moments can be calculated by the following two methods.

\subsection*{Moments by integral}

Some studies have calculated plasma moments by integral in multiple environments, such as the Jupiter's magnetosphere \citep{Mauk2004}, the Martian space \citep{Franz2006}, and the solar wind \citep{Nicolaou2023}.

The integral method starts with moments of the distribution function. The n-order moments of the distribution function $f(\bm{v})$ of a given particle species are defined by the following integral:

\begin{equation}
	\bm{M_n} = \int f(\bm{v}) \bm{v}^n d^3v
\end{equation}

where $\bm{v}^n$ is an n-order tensor, and $d^3v$ is the volume element in velocity space.

Then the number density N, bulk velocity vector $\bm{V}$, pressure tensor $\bm{P}$ can be obtained from zero-, first-, second- order moments respectively:

\begin{equation}
	N = \bm{M_0} = \int f(\bm{v}) d^3v
\end{equation}

\begin{equation}
	N\bm{V}= \bm{M_1} = \int f(\bm{v}) \bm{v} d^3v
\end{equation}

\begin{equation}
	\bm{P} = m\bm{M_2} - \rho \bm{V}\bm{V} = m \int f(\bm{v}) (\bm{v} - \bm{V})(\bm{v} - \bm{V}) d^3v
\end{equation}

And the scalar pressure and temperature can be obtained from  the trace of the associated tensors: $p = tr(\bm{P})/3$ and $T = p/N\kappa$, where $\kappa$ is the Boltzmann constant.

Using ~\cref{c to f} and replacing integral by summation, the above formulas can be rewritten as :

\begin{equation}
	N  = \sum_{E,\theta,\phi} \frac{2J(E,\theta,\phi)}{v} sin\theta \frac{\Delta v}{v}\Delta \theta \Delta \phi
\end{equation}

\begin{equation}
	\bm{V}  = \frac{1}{N} \sum_{E,\theta,\phi} \frac{2J(E,\theta,\phi)}{v} \bm{v} sin\theta \frac{\Delta v}{v}\Delta \theta \Delta \phi
\end{equation}

\begin{equation}
	\bm{P}  = m \sum_{E,\theta,\phi} \frac{2J(E,\theta,\phi)}{v} (\bm{v}-\bm{V})(\bm{v}-\bm{V}) sin\theta \frac{\Delta v}{v}\Delta \theta \Delta \phi
\end{equation}

where $\Delta v, \Delta \theta, \Delta \phi$ are the intervals of $E,\theta,\phi$ channels in the observation.

\subsection*{Moments by fitting} \label{Sec:moments by fitting}
Another way to calculate the moments of a given particle species is to find the best fit of the measured distribution. This method is also widely used in many studies in different regions, such as the Martian space \citep{Fan2022,Franz2006}, the Saturn's magnetosphere \citep{Livi2014}, and the solar wind \citep{Abraham2022,Elliott2016}.  

The fitting method starts with the assumption that the phase space density of a given particle species has a Maxwellian distribution in velocity space\citep{Franz2006,Livi2014}:

\begin{equation}
	f(\bm{v}) = N \big(\frac{m}{2\pi \kappa T_x}\big)^{\frac{1}{2}}\big(\frac{m}{2\pi \kappa T_y}\big)^{\frac{1}{2}}\big(\frac{m}{2\pi \kappa T_z}\big)^{\frac{1}{2}} e^{-\frac{m(v_x-V_x)^2}{2 \kappa T_x}} e^{-\frac{m(v_y-V_y)^2}{2 \kappa T_y}} e^{-\frac{m(v_z-V_z)^2}{2 \kappa T_z}} 
\end{equation}

where $N$ is the number density; $v_i$, $V_i$, $T_i$ is the particle velocity, bulk velocity and temperature in i-th direction, respectively.

\begin{figure}[H]
	\centering
	\includegraphics[width=\linewidth]{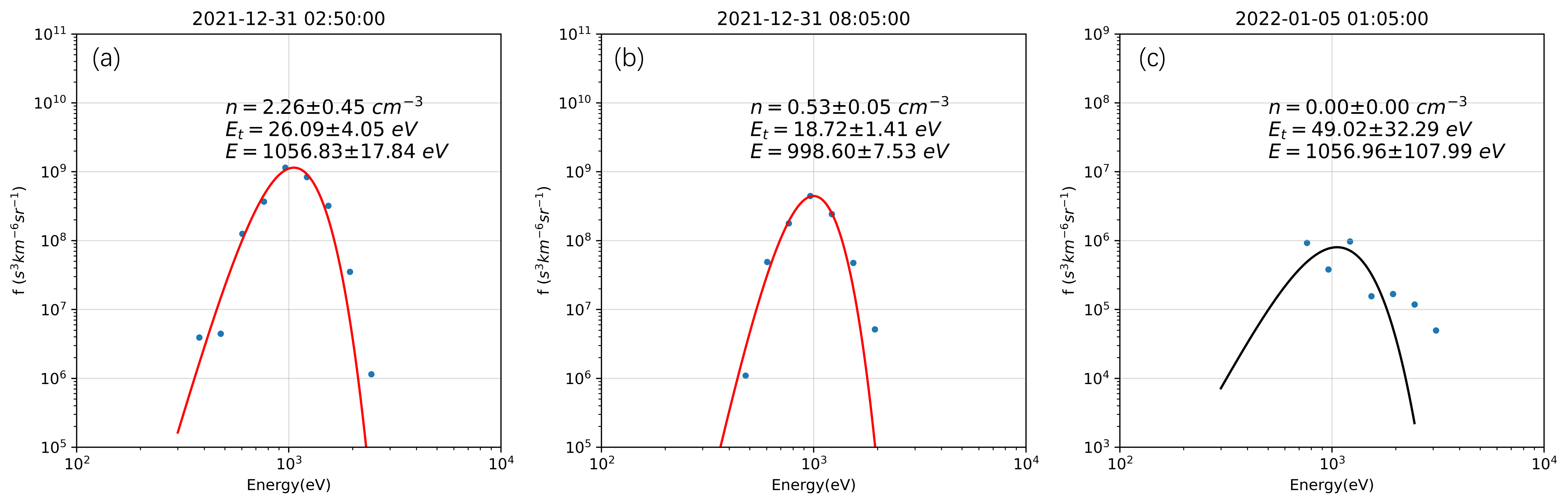}\\
	\caption{The phase space density of $H^+$ calculated from the corrected count data using ~\cref{c to f} during three time periods when Tianwen-1 located in the solar wind. The data are accumulated over 16 seconds (observation time resolution). The red or black lines represent the fitting curves using ~\cref{fitting}. The texts in the figure show the fitting parameters with the standard deviation, including the number density $N$ in $cm^{-3}$, the mean energy $E_m$ and the thermal energy $E_t$ in $eV$. (a \& b) show examples of good fits and (c) is an example of a bad fit due to the limited FOV.}
	\label{fitting_example} 
\end{figure}

Considering that the direction of the ion flows is mainly along the x-axis of the MSO coordinate system, we only choose one direction channel that measures the highest flux in order to observe the flow of the plasma \citep{Livi2014}. Assuming temperature isotropy ($T_x=T_y=T_z=T$) and that the particle velocity, the bulk velocity is mainly along the x-axis. Then the formula can be converted to the following form:

\begin{equation}
	f(v) = N \big(\frac{m}{2\pi \kappa T}\big)^{\frac{3}{2}} e^{-\frac{m(v-V)^2}{2 \kappa T}} 
\end{equation}

where $v=v_x$ and $V=V_x$. Converting velocities to energies we get:

\begin{equation} \label{fitting}
	f(E) =  N\big(\frac{m}{2\pi E_t}\big)^{\frac{3}{2}} e^{-\frac{(\sqrt{E}-\sqrt{E_m})^2}{E_t}}
\end{equation}

where $E_t = \kappa T$, $E_m = \frac{1}{2} m V^2$. By fitting $f(E)$ derived from $J(E, \theta, \phi)$, we can obtain the particle number density $N$, mean energy $E_m$ i.e. bulk velocity $V$, thermal energy $E_t$ i.e. temperature $T$ and thermal pressure by $p=N\kappa T$. To determine whether the fit is good or not, we use the relative deviation of the fit parameters (i.e., $\frac{\Delta N}{N}, \ \frac{\Delta E_m}{E_m}, \ \frac{\Delta E_t}{E_t}$) as a judging metric. The energy interval of the fit is automatically adjusted to minimize the relative deviation, and the parameters are discarded if the final relative deviation is still greater than $30 \%$. ~\cref{fitting_example} show fitting examples using the above equation and metric during three periods when MINPA observes the solar wind. As can be seen in the figure, (a \& b) represent good fits with small relative deviations. While (c) suffers from the limited FOV and shows a bad fit with a high relative deviation, which will be discarded in the final moments data.

\section*{Acknowledgments}

We acknowledge the use of data from the SWIA onboard MAVEN spacecraft, which we obtained from the NASA Planetary Data System (\href{https://pds-ppi.igpp.ucla.edu/}{https://pds-ppi.igpp.ucla.edu/}). One may apply for the Tianwen-1 MINPA and MOMAG data at CNSA Data Release System (\href{https://clpds.bao.ac.cn/web/enmanager/home}{https://clpds.bao.ac.cn/web/enmanager/home}). This work is supported by the NSFC (Grant Nos 42130204, 42188101 and 42241112), the Strategic Priority Program of the Chinese Academy of Sciences (Grant No. XDB41000000) and the Key Research Program of the Chinese Academy of Sciences(ZDBS-SSW-TLC00103). Y.W. is particularly grateful for the support of the Tencent Foundation.

\bibliographystyle{jasr-model5-names}
\bibliography{ref}

\end{document}